\documentclass[review,times,authoryear]{elsarticle}
\usepackage{graphicx}
\usepackage{amssymb}
\usepackage{multirow}
\usepackage{lineno}
\begin{document}

\title {Spin crossover in (Mg,Fe$^{3+}$)(Si,Fe$^{3+}$)O$_3$ bridgmanite: effects of disorder, iron concentration, and temperature}

\author[spa]{Gaurav Shukla}\corref{cor1}
\ead{shuk0053@umn.edu}
\author[spa,cems]{Renata M. Wentzcovitch}
\ead{wentz002@umn.edu}

\cortext[cor1]{Corresponding author}
\address[spa] {School of Physics and Astronomy, University of Minnesota, Minneapolis, Minnesota, USA}
\address[cems]{Department of Chemical Engineering and Materials Science, University of Minnesota, Minneapolis, Minnesota, USA}

\date{\today}
\setlength{\oddsidemargin}{-10pt}

\begin{abstract}

The spin crossover of iron in Fe$^{3+}$-bearing bridgmanite, the most abundant mineral of the Earth’s lower mantle, is by now a well-established phenomenon, though several aspects of this crossover remain unclear. Here we investigate effects of disorder, iron concentration, and temperature on this crossover using \textit{ab initio} LDA + U$_{sc}$ calculations. The effect of concentration and disorder are addressed using complete statistical samplings of coupled substituted  configurations in super-cells containing up to 80 atoms. Vibrational/thermal effects on the crossover are addressed within the quasiharmonic approximation. The effect of disorder seems quite small, while increasing iron concentration results in considerable increase in crossover pressure. Our calculated compression curves for iron-free,  Fe$^{2+}$-, and Fe$^{3+}$-bearing bridgmanite compare well with the latest experimental measurements. The comparison also suggests that in a close system, Fe$^{2+}$ present in the sample may transform into Fe$^{3+}$ by introduction of Mg and O vacancies with increasing pressure. As in the spin crossover in ferropericlase, this crossover in bridgmanite is accompanied by a clear volume reduction and an anomalous softening of the bulk modulus throughout the crossover pressure range. These effects reduce significantly with increasing temperature. Though the concentration of [Fe$^{3+}$]$_{Si}$ in bridgmanite may be small, related elastic anomalies may impact the interpretation of radial and lateral velocity structures of the Earth's lower mantle.
\end{abstract}


\maketitle
 
\section{Introduction}
Bridgmanite (br), (Mg,Fe,Al)(Si,Fe,Al)O$_3$ perovskite (Pv), is the main constituent of the Earth's lower mantle along with (Mg,Fe)O, CaSiO$_3$ Pv, and (Mg,Fe,Al)(Si,Fe,Al)O$_3$ post-perovskite (PPv). Thermodynamics and elastic properties of these minerals provide a direct link to seismic tomographic models. Understanding the effect of iron (Fe) and/or aluminum (Al) substitutions on the physical, chemical, and thermodynamic properties of the host mineral is essential to constrain the composition and thermal structure of the Earth's lower mantle. Ferropericlase (Fp), (Mg,Fe)O, is known to undergo a pressure induced spin crossover from the high (S=2) to the low-spin (S=0) state, which affects its elastic and thermal properties \citep{Badro03,Goncharov,Tsuchiya06,Fei07,Crowhurst,Marquardt09a,Wu09,Wentzcovitch09,Antonangeli11,Mao11,Wu13,Hsu14,Wu14}.

In the case of iron-bearing bridgmanite, in spite of considerable progress of experimental measurements at high pressures and high temperatures \citep{Badro04,Li04,Jackson05,Li06,Lin08,Lundin,McCammon,Dubrovinsky10,McCammon10,Ballaran,Chantel12,Fujino12,Hummer12,Lin12,Dorfman,Lin13,McCammon13,Sinmyo14,Mao15}, deciphering the valence and spin states of multivalent iron and its influence on the physical properties has been quite a formidable challenge due to complexity of the perovskite structure. Iron in bridgmanite may exists in ferrous (Fe$^{2+}$) and ferric (Fe$^{3+}$) states. Fe$^{2+}$ occupies the A-site ([Fe$^{2+}$]$_{Mg}$), while Fe$^{3+}$ can occupy A- ([Fe$^{3+}$]$_{Mg}$) and/or B-site ([Fe$^{3+}$]$_{Si}$) of the perovskite structure  \citep{Badro04,Li04,Jackson05,Li06,Stackhouse07,Bengtson09,Lin08,Dubrovinsky10,Hsu10,Hsu11,Fujino12,Hummer12,Lin12,Lin13,Tsuchiya13,Sinmyo14,Mao15}. In the entire lower mantle pressure-range, [Fe$^{2+}$]$_{Mg}$ remains in the HS state (S=2) but undergoes a pressure induced lateral displacement resulting in the state with increased iron M\"{o}ssbauer quadrupole splitting (QS) \citep{McCammon,Bengtson09,Hsu10,Lin12,Lin13,McCammon13,potapkin13,kupenko14,Shukla15,Shukla15b}. By contrast, [Fe$^{3+}$]$_{Si}$ undergoes a crossover from HS (S=5/2) to LS (S=1/2) state, while [Fe$^{3+}$]$_{Mg}$ remains in the HS (S=5/2) state \citep{Catalli10,Hsu11,Lin12,Lin13,Tsuchiya13,Mao15,Xu15}.

The onset of the HS to LS crossover of [Fe$^{3+}$]$_{Si}$ in Fe$^{3+}$-bearing bridgmanite (Fe$^{3+}$-br) is still much debated. For Fe$^{3+}$-br with 10 mol.\% Fe$_2$O$_3$, \citet{Catalli10} observed  the crossover completion at 48 GPa by X-ray emission spectroscopy(XES), and a change in electronic configuration between 53 and 63 GPa by Synchrotron M$\ddot{o}$ssbauer spectroscopy (SMS). These observations led them to conclude the crossover pressure range to be  approximately 48-63 GPa. Using SMS, \citet{Lin12} found the crossover pressure range 13-24 GPa for a sample containing about $\sim$2.0-2.5$\%$ of Fe$^{3+}$. \citet{Lin12} and \citet{Mao15} further argued that the lower crossover pressure observed by them could be related to the smaller Fe$^{3+}$ concentration in their samples. Using first-principles static LDA + U$_{sc}$ and GGA + U$_{sc}$ calculations for (Mg$_{1-x}$Fe$_x^{3+}$)(Si$_{1-x}$Fe$_x^{3+}$)O$_3$ with $x$ = 0.125, \citet{Hsu11} estimated the crossover pressure 41 GPa and 70 GPa, respectively, while \citet{Tsuchiya13} reported 44 GPa for $x$ = 0.0625 using LDA + U calculations. A thermodynamic model by \citet{Xu15} estimated the Fe$^{3+}$/$\sum$Fe ratio under lower mantle conditions to be $\sim$0.01-0.07 in Al-free bridgmanite. In an effort to understand and reconcile observations and results of these studies, we have investigated the effect of 1) disordered substitution of nearest neighbor Fe$^{3+}$-Fe$^{3+}$ pairs,  2) Fe$^{3+}$ concentration, and 3) vibrational effects on the HS to LS crossover in Fe$^{3+}$-bearing bridgmanite. 

\section{Computational details and Methodology}

\subsection{Computational details}
\begin{figure*}\centering
\includegraphics[width=16cm]{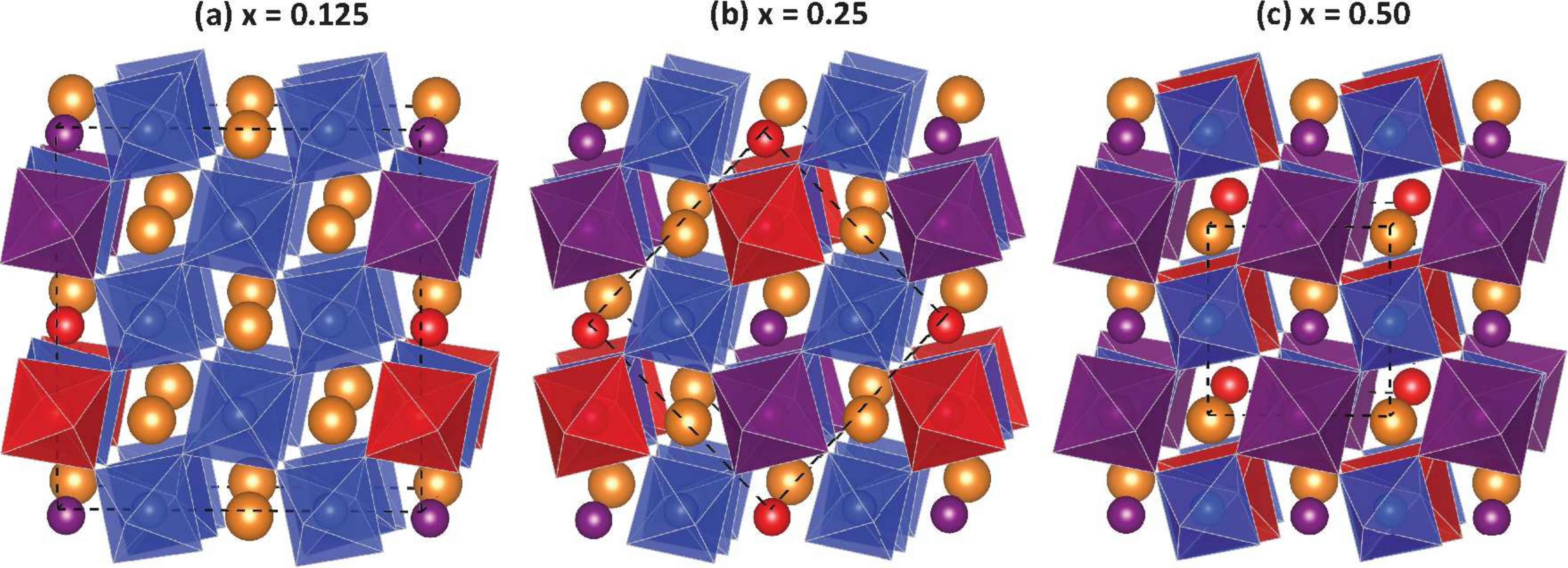}
\caption{(Color online)  Lowest enthalpy atomic configurations for two (purple and red) nearest neighbor [Fe$^{3+}$]$_{Mg}$-[Fe$^{3+}$]$_{Si}$ pairs in (Mg$_{1-x}$Fe$^{3+}_x$)(Si$_{1-x}$Fe$^{3+}_x$)O$_3$ bridgmanite. Unit cells consists of 80 atoms (2 $\times$ 2 $ \times$ 1 super-cell) for x = 0.125, 40 atoms ($\sqrt{2}\times\sqrt{2}\times$ 1 super-cell) for x = 0.25, and 20 atoms for x = 0.50 are shown by black dashed lines. There are 21, 13, and 5 configurations in 80, 40 , and 20 atoms unit-cell, respectively. Si-O octahedron and Mg atom is represented by blue and orange color, respectively.}
\label{fig1}
\end{figure*}

Density functional theory (DFT) within the local density approximation (LDA) \citep{ceperley} has been used in this study. It is well known that standard DFT functionals do not capture strong correlation effects of $3d$ and $4f$ electrons properly. For this reason, standard DFT  is augmented by the the self- and structurally consistent Hubbard U$_{sc}$ (LDA + U$_{sc}$ method) \citep{Cococcioni,Kulik,Hsu09}. U$_{sc}$ values reported by \citet{Hsu11} using these methods have been used here. Disordered substitution of Fe$^{3+}$ in (Mg$_{1-x}$Fe$_x^{3+}$)(Si$_{1-x}$Fe$_x^{3+}$)O$_3$ bridgmanite with varying iron concentration has been investigated in 80- (x = 0.125), 40- (x = 0.25), and 20-atoms (x = 0.50) super-cells (Fig.~\ref{fig1}). Ultrasoft pseudo-potentials \citep{Vanderbilt} have been used for Fe, Si, and O. For Mg, a norm-conserving pseudo-potential, generated by von Barth-Car's method, has been used.  A detailed description of these pseudo-potentials has been reported by \citet{Umemoto08}. The plane-wave kinetic energy and charge density cut-off are 40 Ry and 160 Ry, respectively. For 80-, 40-, and 20-atom super-cells, the electronic states were sampled on a shifted $2\times2\times2$, $4\times4\times4$, and $6\times6\times4$ k-point grid, respectively. Structural optimization at arbitrary pressure has been performed using variable cell-shape damped molecular dynamics \citep{Wentzcovitch91,Wentzcovitch93}. Structures are optimized until the inter-atomic forces are smaller than 10$^{-4}$ $Ry/a.u$.  Vibrational density of states (VDOS) for Fe$^{3+}$ concentration x = 0.125 has been calculated in a 40-atom super-cell using density functional perturbation theory (DFPT) \citep{Baroni} within the LDA + U$_{sc}$ functional \citep{Floris}. For this purpose, dynamical matrices on a $2\times2\times2$ q-point grid of a 40-atom cell were calculated and thus obtained force constants were interpolated on a $8\times8\times8$ q-point grid.  High throughput calculations have been performed using the VLab cyberinfrastructure at the Minnesota Supercomputing Institute \citep{pedro08}.

\subsection{Disordered substitution of Fe$^{3+}$ and spin crossover}\label{disordered_substitution}
Disordered substitution of Fe$^{3+}$ in (Mg$_{1-x}$Fe$^{3+}_{x}$)(Si$_{1-x}$Fe$^{3+}_{x}$)O$_3$ has been studied by replacing nearest neighbor Mg$^{2+}$-Si$^{4+}$ pairs with [Fe$^{3+}$]$_{Mg}$-[Fe$^{3+}$]$_{Si}$ pairs and generating all possible atomic configurations consistent with super-cell size. The number of symmetrically inequivalent configurations, N$_c$, are 21, 13, and 5 in 80- (x = 0.125), 40- (x = 0.25), and 20-atom (x = 0.50) super-cells, respectively. Within the quasiharmonic approximation (QHA) \citep{Umemoto10}, the partition function for disordered system with [Fe$^{3+}$]$_{Si}$ in a spin state $\sigma$ is given by
\begin{eqnarray}
Z^{QHA}_{\sigma}(T,V) &=& \sum_{i=1}^{N_{c}}g_iM_{\sigma} \exp \left\{-\frac{E^{i}_{\sigma}(V)}{k_{B}T}\right\} 
\times \prod^{N_{mode}}_{j=1} \left[\sum_{\nu_{i,j}=0}^{\infty}\exp \left\{ -\left(\nu_{i,j}+\frac{1}{2}\right)\frac{\hbar\omega_{i,j, \sigma}(V)}{k_BT}\right\}\right],
\end{eqnarray}

where $g_{i}$ is the multiplicity of symmetrically equivalent configurations and $E^{i}_{\sigma}$(V) is the static energy of the i$^{th}$ inequivalent configuration at volume $V$. $M_{\sigma}$ is the magnetic degeneracy of the system, which includes the spin and orbital degeneracies. $\nu_{i,j}$ and $\omega_{i,j,\sigma}(V)$  are the number of excited phonons and frequency of j$^{th}$ mode at volume $V$ for i$^{th}$ configuration.  $k_{B}$ and $\hbar$ are Boltzmann and Planck constants, respectively. $N_{c}$ and $N_{mode}$ are the total number of configurations and vibrational modes of a given super-cell. After summing over $\nu_{i,j}$, the partition function is written as
\begin{eqnarray} \label{part-fun}
Z^{QHA}_{\sigma}(T,V) &=& \sum_{i=1}^{N_{c}}g_iM_{\sigma} \exp\left\{-\frac{E^{i}_{\sigma}(V)}{k_{B}T} \right\} 
 \times \prod^{N_{mode}}_{j=1}\left\{\frac{\exp \left( -\frac{\hbar \omega_{i,j,\sigma}(V)}{2k_{B}T}\right)}{1-\exp\left(-\frac{\hbar\omega_{i,j,\sigma}(V)}{k_{B}T}\right)}\right\}.
\end{eqnarray} 

The computation of VDOS (i.e., phonon frequencies $\omega_{i,j,\sigma}(V)$) within DFPT + U$_{sc}$ method \citep{Floris} for every symmetrically inequivalent configuration is extremely challenging. To circumvent this difficulty, we approximate the partition function by assuming that VDOS for a given Fe$^{3+}$ concentration $x$ is same for all configurations. Within this approximation, the partition function becomes
\begin{eqnarray}
 Z^{QHA}_{\sigma}(T,V) &=& \left[\sum_{i=1}^{N_{c}}g_i M_{\sigma}\exp\left\{-\frac{E^{i}_{\sigma}(V)}{k_{B}T}\right\}\right] 
  \times \left[ \prod^{N_{mode}}_{j=1}\frac{\exp\left\{-\frac{\hbar\omega_{j,\sigma}(V)}{2k_{B}T}\right\}}{\left\{1-\exp\left(-\frac{\hbar\omega_{j,\sigma}(V)}{k_{B}T}\right)\right\}}\right]. 
\end{eqnarray}

The Helmholtz free-energy for the system (the super-cell containing $N$ formula units of Fe$^{3+}$-br) with $[Fe^{3+}]_{Si}$ in HS/LS state can be calculated as
\begin{eqnarray} \label{Helmholtz}
F_{HS/LS}(T,V) &=& -k_{B}T\ln \left[Z^{QHA}_{HS/LS}(T,V)\right] \\ \nonumber
 &=& F^{conf}_{HS/LS}(T,V) + F^{vib}_{HS/LS}(T,V) + F^{mag}_{HS/LS}(T,V),
\end{eqnarray}
where $F^{conf}_{HS/LS}(T,V)$ is the free-energy contribution due to the statistical distribution of symmetrically inequivalent configurations and $F^{vib}_{HS/LS}(T,V)$ is the vibrational contribution. The magnetic contribution to the free-energy, $F^{mag}_{HS/LS}(T,V)$, is
\begin{eqnarray}
 F^{mag}_{HS/LS}(T,V) = -k_BTln\left(M_{HS/LS}\right).
\end{eqnarray}
 As mentioned before, Fe$^{3+}$ in (Mg$_{1-x}$Fe$^{3+}_{x}$)(Si$_{1-x}$Fe$^{3+}_{x}$)O$_3$ occupies as a coupled [Fe$^{3+}$]$_{Mg}$-[Fe$^{3+}$]$_{Si}$ pair with molar fraction $x$, where [Fe$^{3+}$]$_{Mg}$ is always in the HS state while [Fe$^{3+}$]$_{Si}$ undergoes HS to LS crossover. Therefore, magnetic degeneracies M$_{HS}$ and M$_{LS}$ for a system containing $N$ formula units of Fe$^{3+}$-br in a super-cell (i.e., super-cell having $Nx$ [Fe$^{3+}$]$_{Mg}$-[Fe$^{3+}$]$_{Si}$ pairs) are given by

\begin{eqnarray}
 M_{HS} &=& \left[m_{HS}(S_{HS}+1)\times m_{HS}(S_{HS}+1)\right]^{Nx} \quad and \\ \nonumber
 M_{LS} &=& \left[m_{HS}(S_{HS}+1)\times m_{LS}(S_{LS}+1)\right]^{Nx}, \quad \quad \quad
\end{eqnarray}
where $m_{HS/LS}$ and $S_{HS/LS}$, respectively, are the orbital degeneracy and total spin of Fe$^{3+}$ in HS and LS state. 
The Gibb's free-energy of the system with $[Fe^{3+}]_{Si}$ in HS/LS state can be calculated as $G_{HS/LS}(T,V) = F_{HS/LS}(T,V) + PV$, which is then converted to $G_{HS/LS}(T,P)$.

To investigate the effect of HS to LS crossover of [Fe$^{3+}$]$_{Si}$, we consider the mixed state (MS) of HS and LS within the ideal solid solution approximation. This solid solution model of HS and LS iron is carried out in the [Fe$^{3+}$]$_{Si}$ sub-lattice only and is very appropriate for this type of problem \citep{Wentzcovitch09, Wu09,Wu13}. The Gibb's free-energy per formula unit for the  MS state is given by
\begin{eqnarray}\label{Gibbs}
 G(P,T,n) &=& nG_{LS}(P,T)+(1-n)G_{HS}(P,T) + G^{mix}(P,T),
\end{eqnarray}
where $n$ is the low-spin fraction, and the mixing free-energy G$^{mix}(P,T)$ is
\begin{eqnarray}
 G^{mix}(P,T) = k_BTx[nln(n) +(1-n)ln(1-n)].
\end{eqnarray}
Minimizing free-energy $G(P,T,n)$ (Eq.~\ref{Gibbs}) with respect to LS fraction, n, we obtain
\begin{eqnarray}\label{n_LS}
 n(P,T)= \frac{1}{1+\frac{m_{HS}(2S_{HS}+1)}{m_{LS}(2S_{LS}+1)}\exp\left\{\frac{\Delta G^{conf+vib}_{LS-HS}}{xk_BT}\right\}},
\end{eqnarray}
where $\Delta G^{conf+vib}_{LS-HS} = G^{conf+vib}_{LS} - G^{conf+vib}_{HS}$. For [Fe$^{3+}$]$_{Si}$ enclosed by ordered oxygen octahedron, orbital degeneracies are m$_{HS}$  = 1 and m$_{LS}$ = 3. Since the degeneracies of e$_g$ and t$_{2g}$ in the perovskite structure (as in the case of Fe$^{3+}$-bearing bridgmanite) are broken due to the presence of asymmetry in oxygen octahedron, m$_{LS}$ = $1$ has been used in most of the cases. However, to assess the effect of orbital degeneracies, m$_{LS}$ = $3$ case has also been tested while addressing the vibrational effects.

\section{Results and discussion}
\subsection{Effect of disorder on spin crossover}
\begin{figure*}\centering
\includegraphics [width=16.cm] {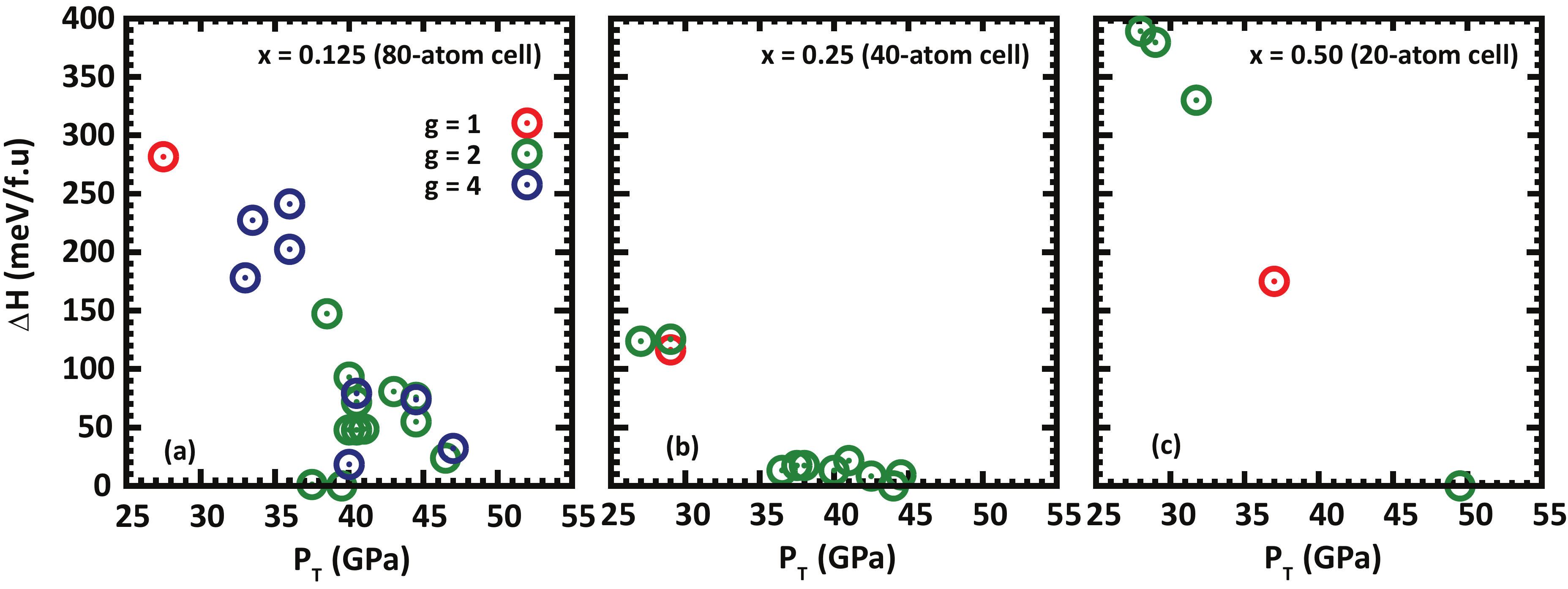}
\caption{(Color online) Relative enthalpy per formula unit (f.u) at crossover pressures of all configurations with respect to lowest enthalpy configuration for (a) $x = 0.125$ (80 atoms), (b) $x = 0.25$ (40 atoms), and (c) $x = 0.50$ (20 atoms), respectively. g is multiplicity of configurations shown by red, green, and blue colors.}
\label{fig2}
\end{figure*}

\begin{figure*}\centering\centering
\includegraphics[width=8cm]{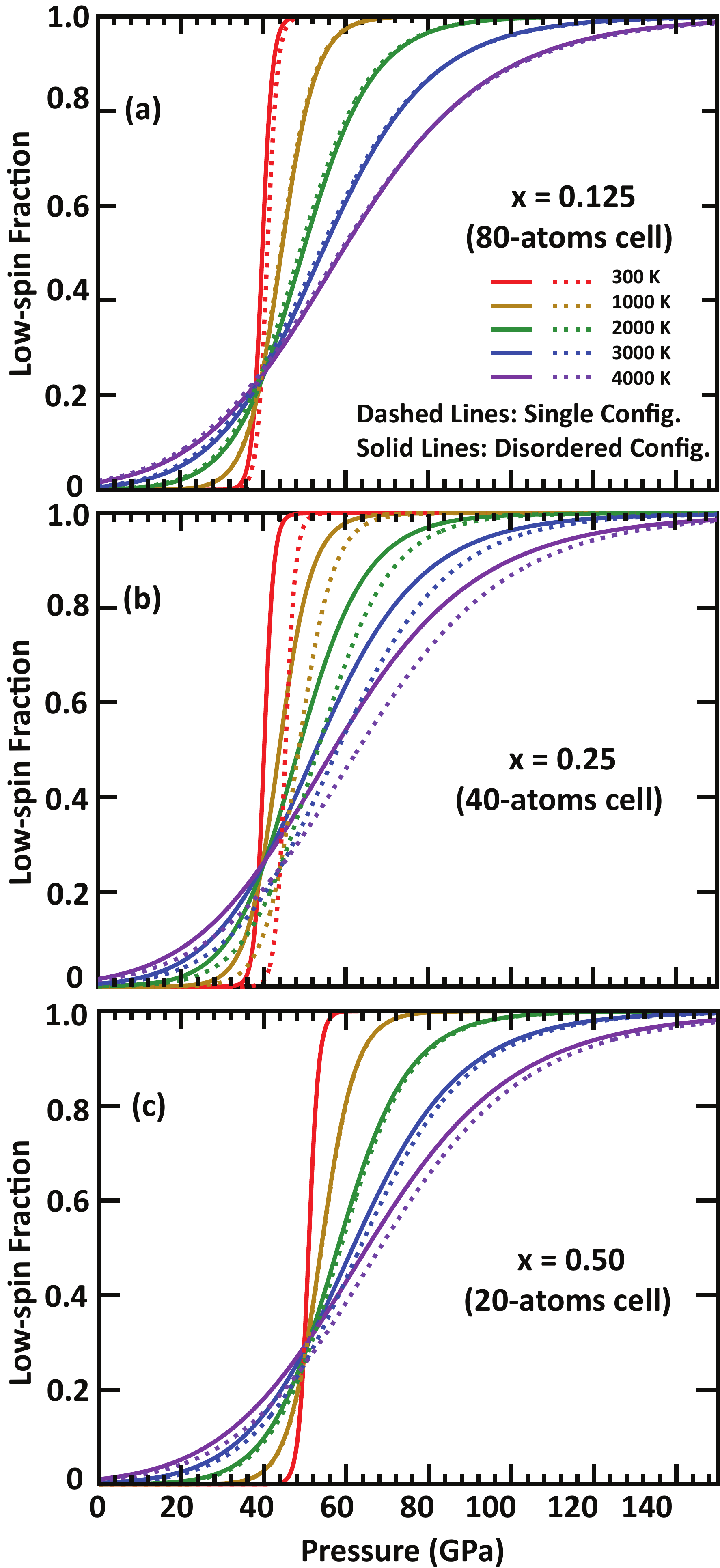}
\caption{(Color online) Pressure and temperature dependence of low-spin fraction (n) of [Fe$^{3+}$]$_{Si}$  in (Mg,Fe$^{3+}$)(Si,Fe$^{3+}$)O$_3$ bridgmanite in a) 80 , (b) 40, and (c) 20 atom super-cells, respectively. Solid lines represent disordered configurations while dashed lines are for the lowest enthalpy configuration.}
\label{fig3}
\end{figure*}

We investigate the solid solution of (Mg$_{1-x}$Fe$_x$)(Si$_{1-x}$Fe$_x$)O$_3$ by sampling the complete phase space of configurations of [Fe$^{3+}$]$_{Mg}$-[Fe$^{3+}$]$_{Si}$ pairs in super-cells containing 80-, 40-, and 20-atoms. We consider only configurations with coupled substitution, i.e., [Fe$^{3+}$]$_{Mg}$-[Fe$^{3+}$]$_{Si}$ pairs since they are the lowest energy configurations \citep{Stackhouse07,Hsu11}. Representative crystal structures for this type of substitutions are shown in Fig.~\ref{fig1}. Two [Fe$^{3+}$]$_{Mg}$-[Fe$^{3+}$]$_{Si}$ pairs (shown in purple and red color) are substituted in the 80, 40, and 20-atoms super-cell for x = 0.125, 0.25, and 0.50, respectively. There are 21, 13, and 5 symmetrically inequivalent configurations, respectively, in these super-cells. Static crossover pressures in [Fe$^{3+}$]$_{Si}$  and their relative enthalpy at crossover point for each configurations are shown in Fig.~\ref{fig2}.  Crossover pressures for these configurations are in $\sim$27.5 GPa to $\sim$50.0 GPa pressure range. The lowest enthalpy configurations are shown in Fig.~\ref{fig1}. In these configurations, Fe$^{3+}$ ions are ordered in (010) plane of the perovskite structure. The trend observed here is similar to the one noticed by \citep{Umemoto08} for Fe$^{2+}$ substitution in bridgmanite, where Fe$^{2+}$ ions preferentially order in the (110)-plane. This observation suggests that Fe$^{3+}$ substitution in bridgmanite prefers to cluster at lower mantle pressure conditions. For better understanding, this ordering effect should be investigated by sampling all possible configurations in very large super-cells and performing solid solution calculations. However, for practical reason we must limit ourselves to smaller super-cells here.

First, we investigate the effect of disordered substitution of Fe$^{3+}$ in bridgmanite. For this purpose, the free-energy for HS/LS states of [Fe$^{3+}$]$_{Si}$ has been calculated using Eq.~\ref{Helmholtz} by disregarding the vibrational contribution. The LS fraction, $n(P,T)$, shown in Fig.~\ref{fig3}, was calculated using Eq.~\ref{n_LS}. $n(P,T)$ in the disordered system depends on the number of configurations (N$_{c}$) and their enthalpies. For x = 0.125, $n(P,T)$ in the disordered system (solid lines) at 300 K shifts towards the smaller pressure region compared to that of the lowest enthalpy configuration (dashed lines). This is caused by the contribution of the second lowest enthalpy configuration with smaller static crossover pressure (P$_T$) (Fig.~\ref{fig2}a). At higher temperatures, the contribution from other configurations with higher enthalpy and higher P$_T$ shifts the disordered system's $n(P,T)$ towards higher pressures. The number of symmetrically inequivalent configurations decreases drastically with decreasing super-cell size. For x = 0.25 (40-atom super-cell), the lowest enthalpy configuration has the second highest P$_T$ (Fig.~\ref{fig2}b). $n(P,T)$ for the disordered system is shifted to lower pressures due to contributions of configurations with smaller P$_T$ and enthalpy difference below $\sim$25meV. In the case of x = 0.50 (20-atom super-cell), all configurations other than the lowest enthalpy one have much higher enthalpy and do not contribute significantly to change $n(P,T)$ of disordered system at lower temperatures. However, at higher temperatures ($\geq$ 2000 K) other configurations contribute to shift $n(P,T)$ towards smaller pressures (Fig.~\ref{fig2}c). In overall, the crossover pressure increases significantly with increasing Fe$^{3+}$ concentration, which is consistent with previous observations \citep{Lin12,Mao15}.

\subsection{Effect of vibrations}
\begin{figure*}\centering
\includegraphics[width=8cm]{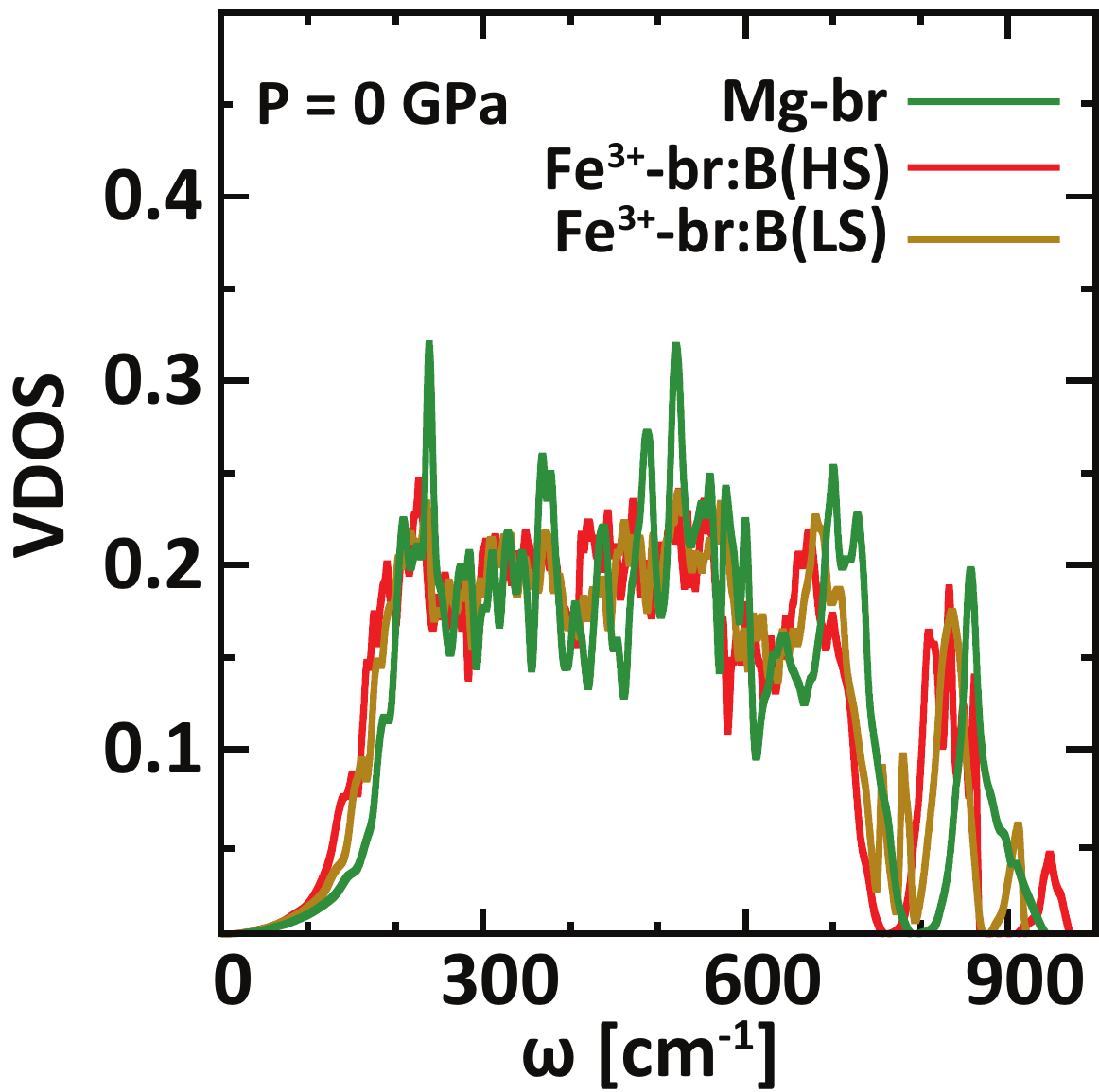}
\caption{(Color online) Vibrational density of states (VDOS) at $0$ GPa for MgSiO$_3$ bridgmanite (Mg-br) using DFPT method \citep{Baroni}, and for HS and LS state states of [Fe$^{3+}$]$_{Si}$ (i.e, ferric iron in B-site of the perovskite structure) in Fe$^{3+}$-bearing bridgmanite (Fe$^{3+}$-br) using DFPT + U$_{sc}$ method \citep{Floris}.}
\label{fig4}
\end{figure*}

Having investigated the effect of disorder, we now focus on vibrational effects. Owing to extremely high computational cost of vibrational density of states (VDOS) calculation) using LDA + U$_{sc}$ functional, we have investigated vibrational effects only for the x = 0.125 system. VDOS calculation for HS/LS states of [Fe$^{3+}$]$_{Si}$ was performed in a 40-atoms super-cell. Example of these VDOS at 0 GPa are shown in Fig.~\ref{fig4}. The VDOS spectrum for Fe$^{3+}$-br shifts towards lower frequencies with respect to that of MgSiO$_3$-bridgmanite (Mg-br) due to increased molecular weight of Fe$^{3+}$-br. The high frequency region of VDOS for the LS state of [Fe$^{3+}$]$_{Si}$ further shifts towards low frequency due to HS to LS crossover. 

\begin{figure*}\centering
\includegraphics[width=14.5cm]{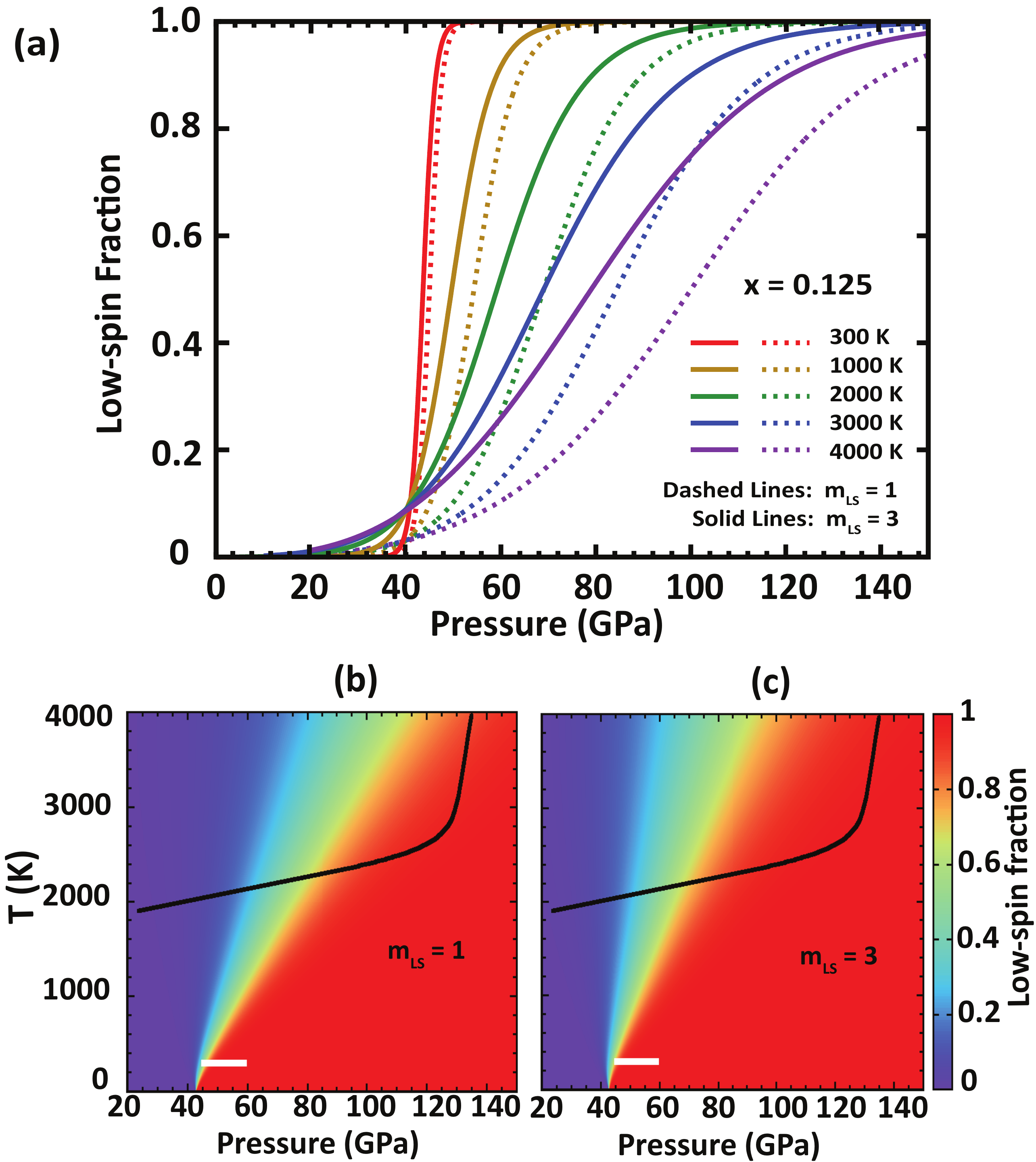}
\caption{(Color online) (a) Low-spin fraction ($n$) of [Fe$^{3+}$]$_{Si}$ in (Mg$_{0.875}$Fe$_{0.125}$)(Si$_{0.875}$Fe$_{0.125}$)O$_3$ bridgmanite when vibrational effects are incorporated using quasiharmonic approximation \citep{Carrier07}. Dashed and solid lines represent the calculations for m$_{LS}$ = 1 and 3, respectively. Pressure temperature phase diagram for HS to LS crossover of [Fe$^{3+}$]$_{Si}$ (b) with m$_{LS}$ = 1, and (c) m$_{LS}$ = 3. Solid black lines in (b and c) represents the lower mantle model geotherm by \citep{Boehler00}, while white bar represents the experimental pressure range at room temperature in which HS to LS crossover of [Fe$^{3+}$]$_{Si}$ is completed \citep{Catalli10}.}
\label{fig5}
\end{figure*} 

The vibrational contribution to the free-energy was calculated within the quasiharmonic approximation (QHA) \citep{Carrier07, Wallace72}. The pressure and temperature dependence of calculated low-spin fraction, $n(P,T)$, is shown in Fig.~\ref{fig5}. The orbital degeneracy for the LS state, m$_{LS}$, is more likely to be 1 due to octahedron asymmetry. However, in order to assess the effect of orbital degeneracy variation,  calculated $n(P,T)$ for m$_{LS}$ = 1 (dashed lines) are also compared with that for m$_{LS}$ = 3 (solid lines). Inclusion of the vibrational contribution to free-energy increases the crossover pressure significantly.  $n(P,T)$ for m$_{LS}$ = 1 and 3 are also shown in Fig.~\ref{fig5}(b) and in Fig.~\ref{fig5}(c), respectively. In overall, $n(P,T)$ for m$_{LS}$ = 1 and 3  are similar except for the difference in Clapeyron slopes (i.e., $dP_T/dT$ for n = 0.5). Since the high temperature crossover broadening is not affected by the choice of m$_{LS}$ values, we will continue this analysis using m$_{LS}$ = 1. The spin crossover of [Fe$^{3+}$]$_{Si}$ shown here is much broader than that reported by \citet{Tsuchiya13}. Our estimated crossover pressure width at 300 K is about $\sim$8 GPa, which agrees fairly well with the experimental measurements \citep{Catalli10,Lin12,Mao15}, while the one reported by \citet{Tsuchiya13} is $<$2 GPa. These differences become more prominent at higher temperatures and may be related to the use of different values of Hubbard U and different techniques for VDOS calculations. The values of U used here were calculated self-consistently \citep{Cococcioni} (i.e., starting form a trial LDA + U ground state, self-consistent U$_{sc}$'s are obtained iteratively), while those reported by \citet{Tsuchiya13} were calculated from LDA  ground states. \citet{Tsuchiya13} used a finite displacement method \citep{Alfe09} to obtain their VDOS and disregarded the calculation of dielectric constant tensor that leads to LO-TO splitting for polar materials.  In this work, we have used DFPT + U$_{sc}$ method developed by \citet{Floris} for VDOS computation. Although computationally expensive, this method is a more precise approach for lattice dynamical calculations. This method has been  applied successfully to address the lateral displacement of iron (a very delicate and highly debated phenomenon) and its associated change of M\"{o}ssbauer quadrupole splitting (QS) in Fe$^{2+}$-bearing bridgmanite \citep{Shukla15}. Therefore, we believe that calculation of VDOS and vibrational free-energy contribution using DFPT + U$_{sc}$ method is quite robust.    

\section{Effect of spin crossover on volume and bulk modulus}
\subsection{Theoretical predictions}
\begin{figure*}\centering
\includegraphics[width=8cm]{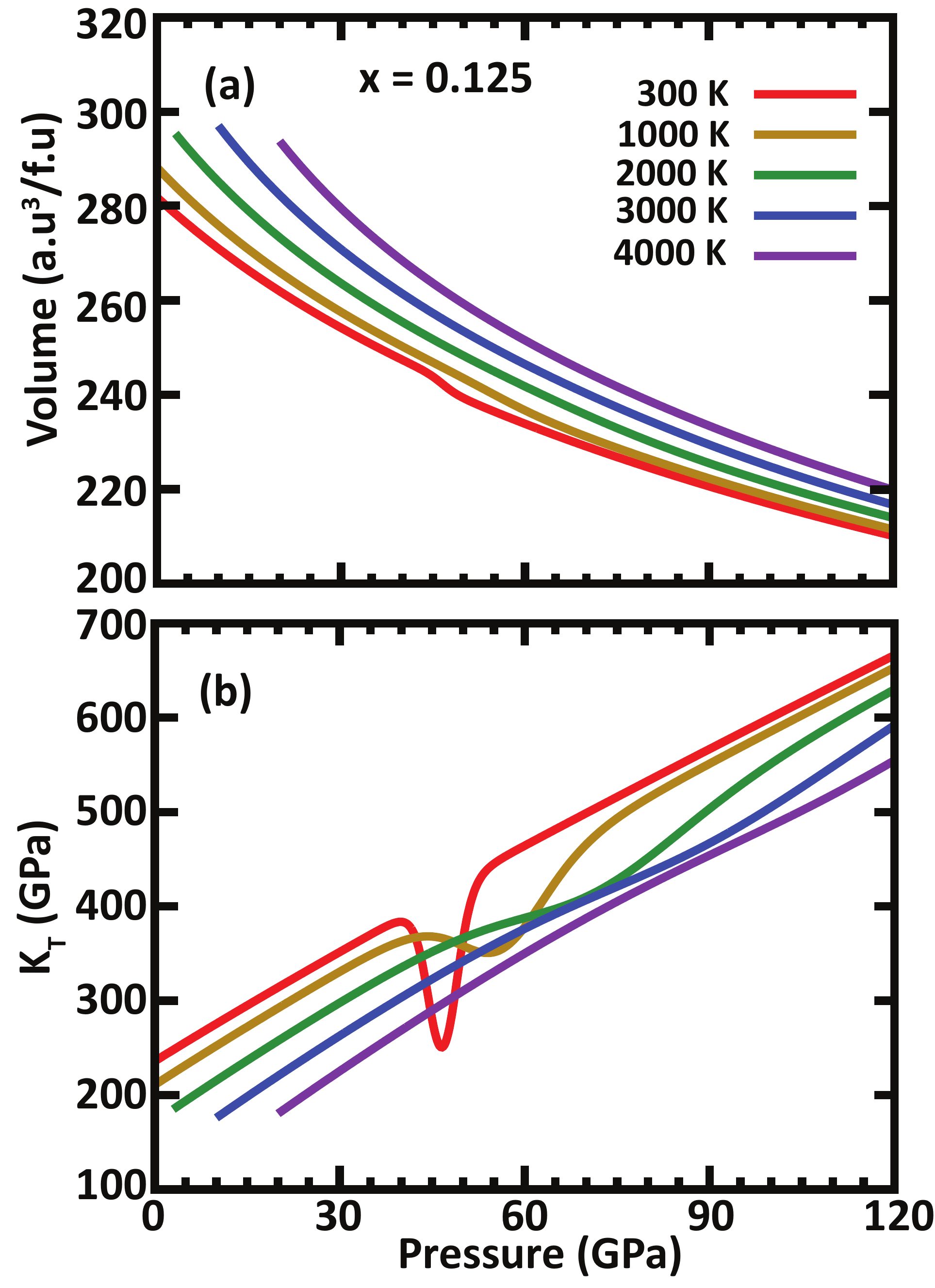}
\caption{(Color online) Pressure and temperature dependence of (a) volume (per formula unit), and (b) isothermal bulk moduli for (Mg$_{0.875}$Fe$_{0.125}$)(Si$_{0.875}$Fe$_{0.125}$)O$_3$ bridgmanite.}
\label{fig6}
\end{figure*}

Spin crossover of [Fe$^{3+}$]$_{Si}$ in Fe$^{3+}$-br is quite sharp in the lower temperature region and broadens significantly at higher temperature (Fig.~\ref{fig5}). To understand the consequences of this spin crossover on elastic properties, we used the solid solution formalism adopted by \citet{Wentzcovitch09} to calculate volume and bulk modulus of the mixed spin state (MS):
\begin{eqnarray}
V_{MS}(T,P) = (1-n)V_{HS}(T,P) + nV_{LS}(T,P),
\end{eqnarray}
where n is the LS fraction.  Using $K_{T}(T,P) = -V_{MS}(dP/dV_{MS})$, isothermal bulk modulus of MS state is given by  
\begin{eqnarray}
\frac{V_{MS}}{K_{T}}=n\frac{V_{LS}}{K^{LS}_T}+(1-n)\frac{V_{HS}}{K^{HS}_T} \nonumber \\
 \left. - (V_{LS} - V_{HS})\frac{\partial n}{\partial P} \right|_T, 
\end{eqnarray}
where $K^{HS/LS}_T$ are the isothermal bulk moduli of HS/LS states. 

The HS to LS crossover of [Fe$^{3+}$]$_{Si}$ goes through volume reduction, which produces anomalous softening in the bulk modulus (K) in the crossover region. The strength of the anomaly depends on  temperature, volume difference ($\Delta V^{HS\rightarrow LS} = V_{LS} - V_{HS}$), and Gibb's free-energy difference ($\Delta G^{HS\rightarrow LS} = G_{LS} - G_{HS}$). For $x$ = 0.01,  $\Delta V^{HS\rightarrow LS}$ is approximately $\sim$-0.15\%, which agrees well with results of previous first-principles calculations \citep{Hsu11,Tsuchiya13} and is fairly comparable to the experimental value of $\sim$-0.2\% \citep{Mao15}. This volume reduction produces a significant bulk modulus softening ($\sim$12\%) at 300 K, which is smeared out with increasing temperature due to broadening of the spin crossover region (Fig.~\ref{fig6}b). In spite of the clear volume reduction throughout the spin crossover, \citet{Tsuchiya13} disregarded the anomalous softening of bulk modulus probably due to very sharp crossover observed in their calculations. However, owing to the considerably broad HS to LS crossover and noticeable volume reduction, as evidenced by experimental measurements \citep{Mao15}, this spin crossover should also be accompanied by a bulk modulus softening anomaly.  

\subsection{Comparison with experiments}
\begin{figure*}\centering
\includegraphics[width=8cm]{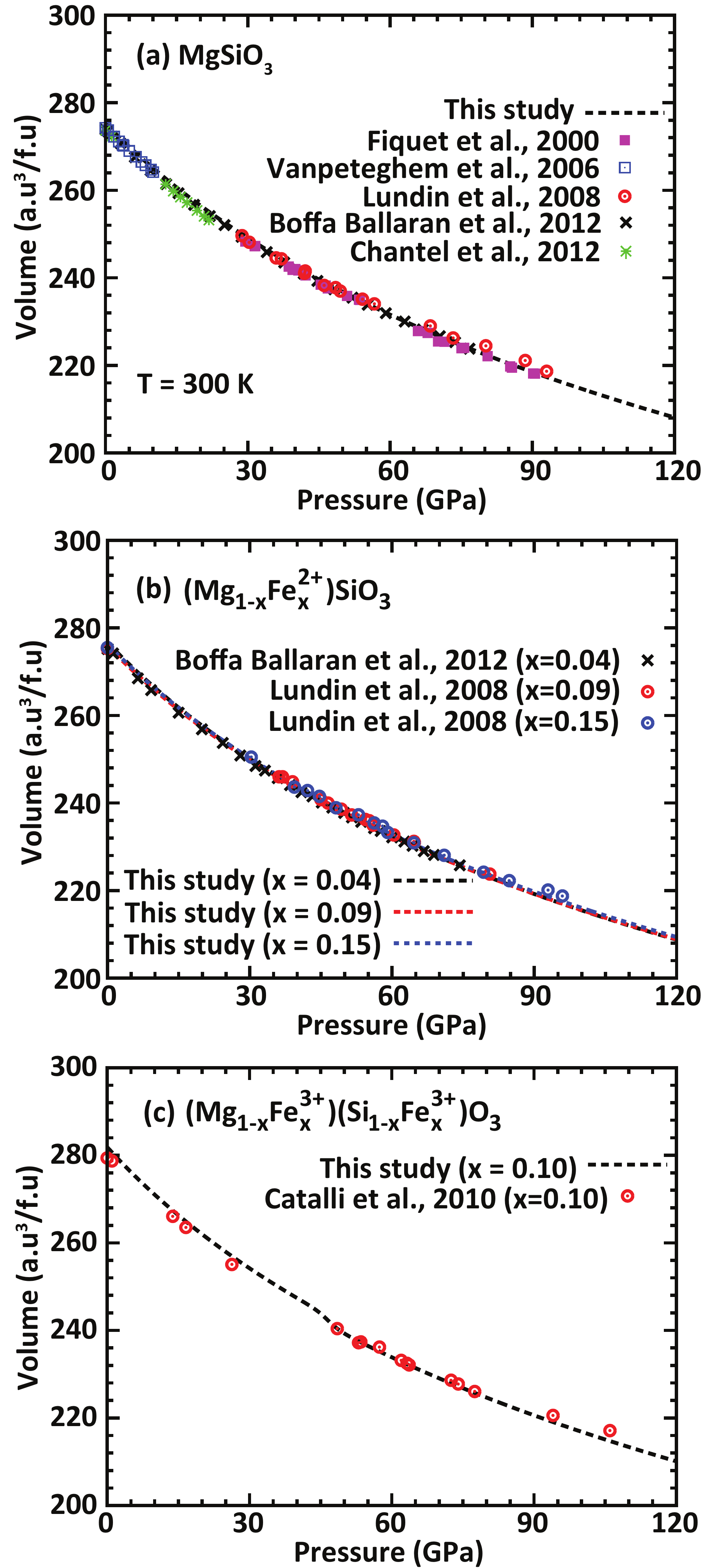}
\caption{(Color online) 300 K compression curves for (a) MgSiO$_3$, (b) (Mg$_{1-x}$Fe$^{2+}_{x}$)SiO$_3$, and (c) (Mg$_{1-x}$Fe$^{3+}_{x}$)(Si$_{1-x}$Fe$^{3+}_{x}$)O$_3$. Our first-principles calculations (lines) are compared with experimental measurements (symbols).}
\label{fig7}
\end{figure*}

Having calculated the compression curves for Fe$^{3+}$-br, we compare our results with the available experimental measurements. Our calculated compression curves for pure MgSiO$_3$, (Mg$_{1-x}$Fe$^{2+}_{x}$)SiO$_3$, and (Mg$_{1-x}$Fe$^{3+}_{x}$)(Si$_{1-x}$Fe$^{3+}_{x}$)O$_3$, shown in Fig.~\ref{fig7}a, \ref{fig7}b, and \ref{fig7}c, respectively, agree well with measurements \citep{Fiquet00,Vanpeteghem06,Lundin,Catalli10,Ballaran,Chantel12}. Compression curves for 0$<x<$0.125 are linearly interpolated using $x$ = 0 and 0.125 results. Using enstatite powder, (Mg$_{0.9}$Fe$_{0.1}$)SiO$_3$, bridgmanite (Br10) was synthesized by \citet{Lin12} and it was suggested that sample may have Fe$^{3+}$/$\sum Fe$ $\approx$ 20\% in the octahedral sites (Si-sites) and Fe$^{2+}$/$\sum Fe$ $\approx$ 80\% in the Mg-sites. Compression curve for this sample at 300 K was recently obtained by \citet{Mao15}. 

In order to gradually transform Fe$^{2+}$ to Fe$^{3+}$ in the sample under pressure, the system should at least be able to 1) incorporate MgO and O$_2$, 2) produce metallic iron, 3) create vacancies, or 4) a combination of these \citep{Xu15}.  \citet{Lin12} synthesis seems to be in a closed system, where Fe$^{3+}$ production from Fe$^{2+}$ is more likely to be accompanied by Mg and O vacancies.  Therefore to compare with measurements by \citet{Mao15}, we modeled their Br10 sample as
 
\begin{eqnarray}\label{Br10}
 Br10 = (Mg_{0.9}Fe_{0.1})SiO_3 &=& a(Mg_{1-x}Fe^{2+}_{x})SiO_{(3)} \nonumber \\
 &+&b(Mg_{1-y -\alpha}Fe^{3+}_{y})(Si_{1-y}Fe^{3+}_{y})O_{3(1-\beta)},
\end{eqnarray}
where stoichiometric coefficients are constrained by $ax + 2by = 0.1$, $\alpha = 2y$, and $\beta = y$. (see Appendix A). 

\begin{figure*}\centering
\includegraphics[width=8cm]{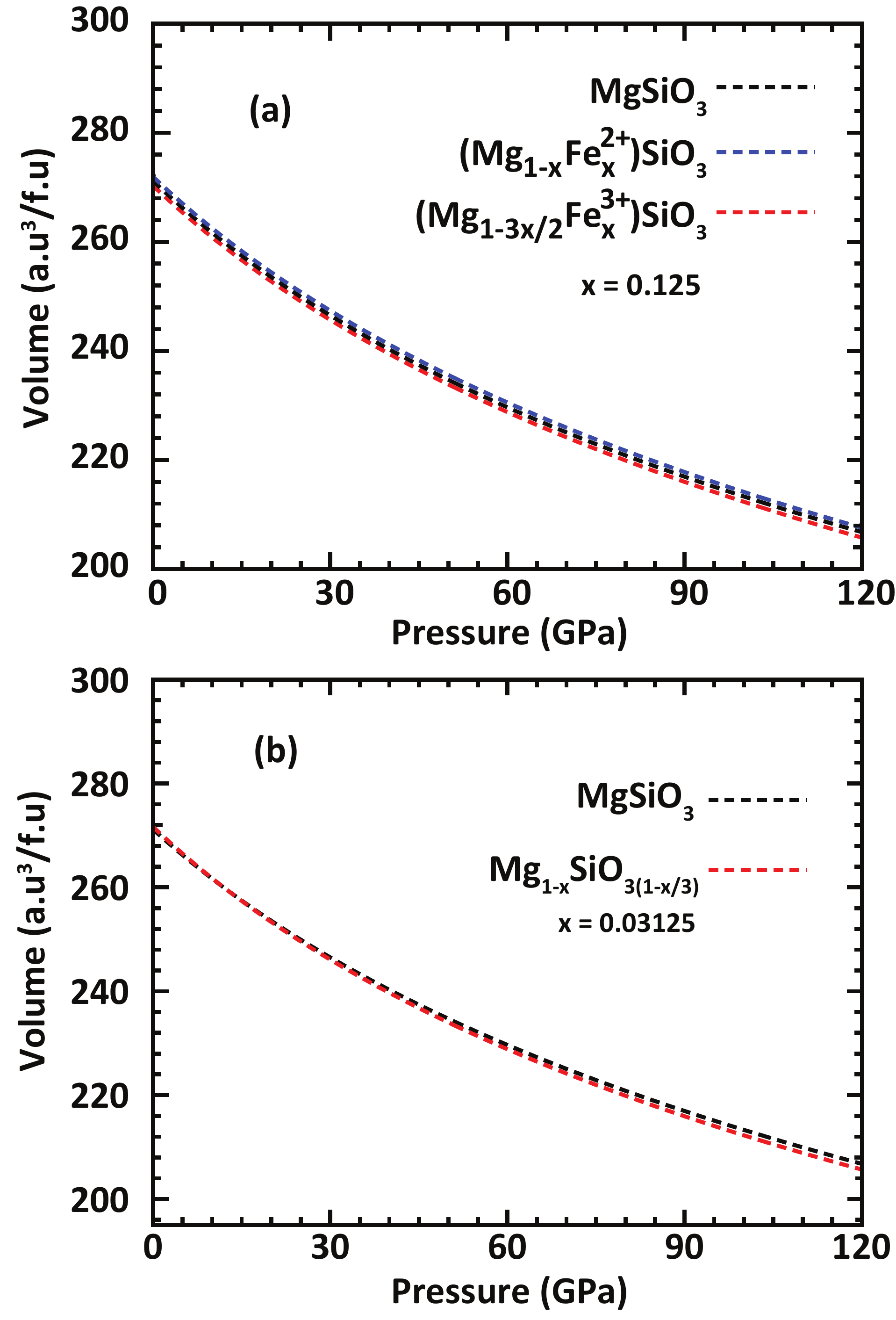}
\caption{(Color online) Static compression curves comparison of (a) MgSiO$_3$ and (Mg$_{1-3x/2}$Fe$^{3+}_{x}$)SiO$_3$ ($x$ = 0.125) and (b) MgSiO$_3$ and Mg$_{1-x}$SiO$_{3(1-x/3)}$ ($x$ = 0.03125).}
\label{fig8}
\end{figure*}

\begin{figure*}\centering
\includegraphics[width=8cm]{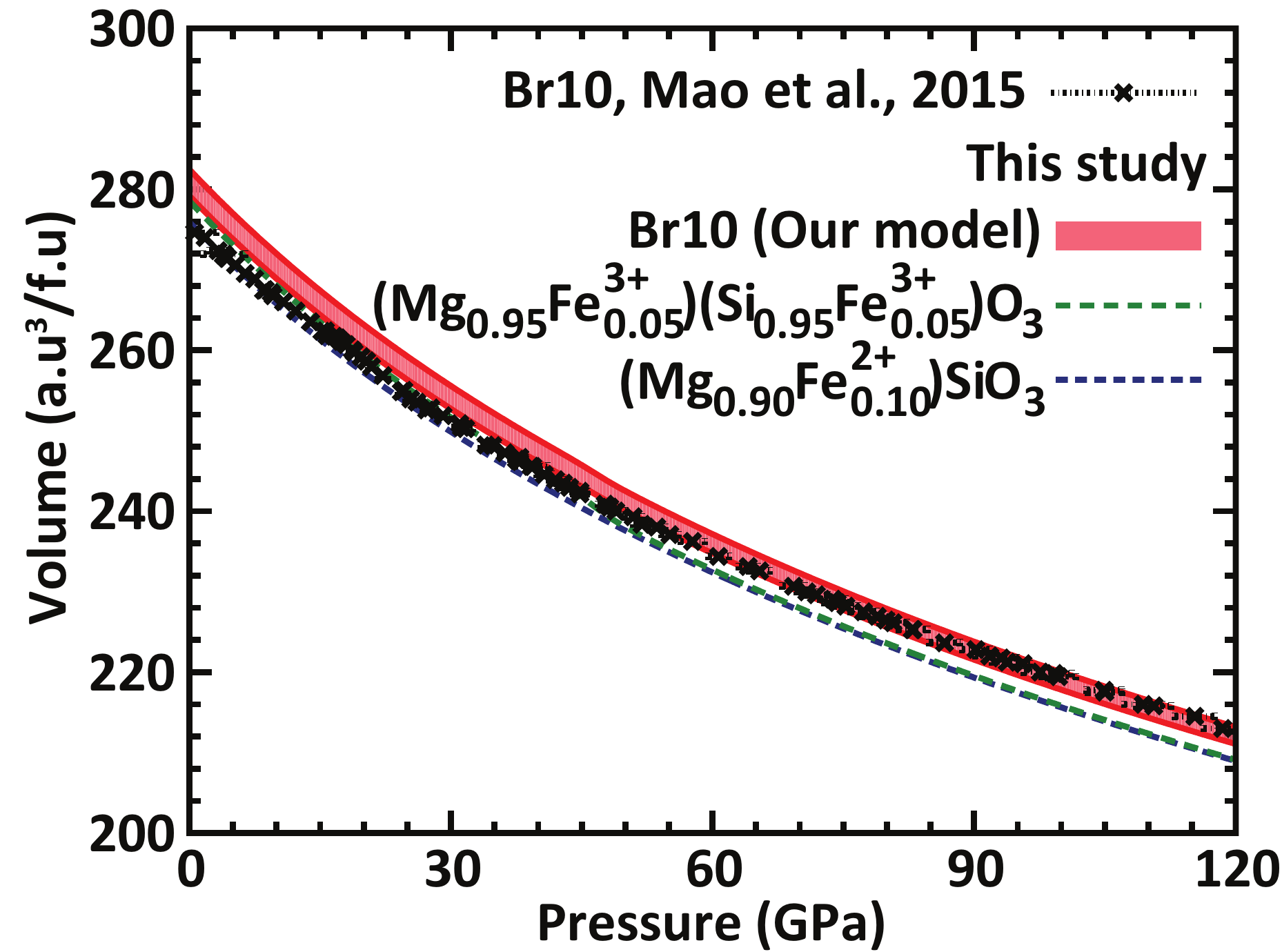}
\caption{(Color online) Compression curve for bridgmanite with 10 mol.\% Fe (Br10). Our first-principles calculations (lines and red filed-curve) are compared with experimental measurements (black symbols). Br10 sample was synthesized by \citet{Lin12} and compression curve was measured by \citet{Mao15}. Here, Br10  sample has been been modeled by balancing the stoichiometry with appropriate molar fractions of (Mg$_{1-x}$Fe$^{2+}_{x}$)SiO$_3$ and (Mg$_{1-x}$Fe$^{3+}_{x}$)(Si$_{1-x}$Fe$^{3+}_{x}$)O$_3$ bridgmanite (red filed-curve). Lower- and upper-bounds of the compression curve are for 10\% and 20\% [Fe$^{3+}$]$_{Si}$/$\sum$Fe in the sample, respectively.}
\label{fig9}
\end{figure*}

These products containing $Mg$ and $O$ vacancies ([V]$_{Mg}$ and [V]$_{O}$) should be created in the absence of other phases in a closed system. It is important, therefore, to investigate the effect of vacancies on compression curves. Fig.~\ref{fig8}(a) compares the static compression curves for pure MgSiO$_3$, (Mg$_{1-x}$Fe$^{2+}_{x}$)SiO$_3$, and (Mg$_{1-3x/2}$Fe$^{3+}_{x}$)SiO$_3$. Fe$^{3+}$ in (Mg$_{1-3x/2}$Fe$^{3+}_{x}$)SiO$_3$ is accommodated exclusively in Mg-site via Mg vacancies (3Mg$^{2+}$ $\rightarrow$ 2[Fe$^{3+}$]$_{Mg}$ + [V]$_{Mg}$). The presence of [Fe$^{2+}$]$_{Mg}$ and [Fe$^{3+}$]$_{Mg}$ in the perovskite A-site alone do not affect the compression curve significantly. The dominant changes in the compression curve are due to [Fe$^{3+}$]$_{Si}$. Therefore, the compression curve for our model Br10 (Eq.~\ref{Br10}), where Fe$^{3+}$ is accommodated via coupled substitution [Fe$^{3+}$]$_{Mg}$-[Fe$^{3+}$]$_{Si}$ with Mg and O vacancies, would be a reasonably close to the one obtained by \citet{Mao15}.  We calculated the compression curves for our model Br10 with 10\%$\le$[Fe$^{3+}$]$_{Si}$/$\sum$Fe$\le$20\%. Stoichiometric coefficients in Eq.~\ref{Br10} for 10\% [Fe$^{3+}$]$_{Si}$/$\sum$Fe (lower-bound) are: a =  b = 0.505, x = 0.15842, y = 0.019802, and for 20\% [Fe$^{3+}$]$_{Si}$/$\sum$Fe (upper-bound) are: a =  b = 0.51, x = 0.11765, y = 0.039216. Details of these stoichiometry calculations are shown in Appendix A. It is worth mentioning here that the total iron concentration in our B10 model is fixed and any variation in [Fe$^{3+}$]$_{Si}$ will be constrained by simultaneous changes in [Fe$^{3+}]_{Mg}$ and [Fe$^{2+}]_{Mg}$. As shown in Fig.~\ref{fig8}(b), the compression curve of Mg$_{1-x}$SiO$_{3(1-x/3)}$ with $x$ = 0.03125 (i.e., MgSiO$_3$ with small amount of Mg and O vacancies) is very similar to that of pure MgSiO$_3$. Therefore, we use  
\begin{eqnarray}
 Br10 = (Mg_{0.9}Fe_{0.1})SiO_3 & \approx & a(Mg_{1-x}Fe^{2+}_{x})SiO_{3} \nonumber \\
 &+&b(Mg_{1-y}Fe^{3+}_{y})(Si_{1-y}Fe^{3+}_{y})O_{3},
\end{eqnarray}
as approximate description of Eq.~(\ref{Br10}). Calculated 300 K compression curve for model Br10 is depicted by red curve in Fig.~\ref{fig9}. \citet{Mao15} data for compression curve (black symbols) tends to agree better with the calculated values for (Mg$_{0.9}$Fe$^{2+}_{0.1}$)SiO$_3$ in the lower-pressure range, while in the higher-pressure range agreement with our model Br10 (red curve) is better. This observation suggests that Fe$^{2+}$ present in the sample at ambient condition may tend to transform into Fe$^{3+}$ with increasing pressure. 

Two clarifications are in order: first, the larger volume of Br10 beyond approximately 45 GPa, compared to that of (Mg$_{0.9}$Fe$^{2+}_{0.1}$)SiO$_3$  or (Mg$_{0.95}$Fe$^{3+}_{0.05}$)Si$_{0.95}$Fe$^{3+}_{0.05}$O$_3$ is achieved by the introduction of vacancies while maintaining the number of moles of each element invariant; second, such conclusion is achieved exclusively on the basis of the compression curve. No energetics has been investigated yet, the main reason being the nature of these vacancies is unclear. There are multiple possibilities, including Mg-O vacancy ordering associated with Fe$^{3+}$. This is not uncommon in transition metal oxides \citep{Schulz93,Leung96} where the metal has multiple valences, iron in this case. Theoretical investigation of this problem is highly desirable but a non-trivial task that goes beyond the scope of this work.    

\section{Geophysical Significance}
\begin{figure*}\centering
\includegraphics[width=8cm]{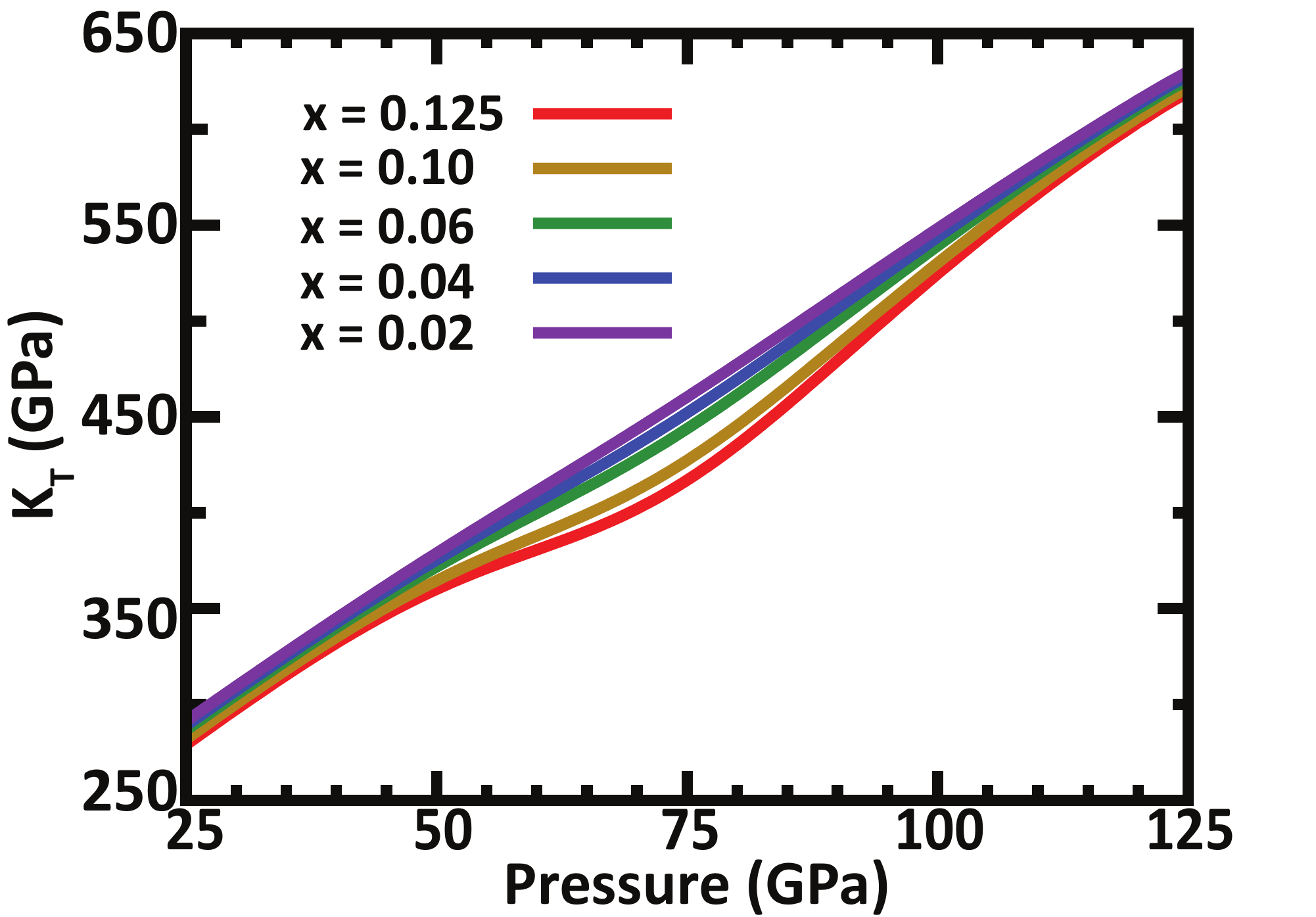}
\caption{(Color online) Bulk moduli (K$_T$) for (Mg$_{1-x}$Fe$^{3+}_x$)(Si$_{1-x}$Fe$^{3+}_x$)O$_3$ bridgmanite along the Earth's lower mantle geotherm \citep{Boehler00} for varying Fe$^{3+}$ concentration.}
\label{fig10}
\end{figure*}

The bulk modulus of (Mg$_{1-x}$Fe$^{3+}_x$)(Si$_{1-x}$Fe$^{3+}_x$)O$_3$ along a lower mantle model geotherm \citep{Boehler00} (black lines in Fig.~\ref{fig5}b and \ref{fig5}c) for several Fe$^{3+}$ concentrations are shown in Fig.~\ref{fig10}. For x = 0.125, the bulk modulus softening for Fe$^{3+}$-br ($\sim$7$\%$) is smaller than that for (Mg,Fe)O ($\sim$11$\%$) \citep{Wu13}. This anomaly reduces rapidly with decreasing $x$ and almost disappears for $x$ = 0.02 concentration. Based on a thermodynamics model, \citet{Xu15} estimated the amount of Fe$^{3+}$/$\sum$Fe to be very small ($\sim$0.01-0.07) in Al-free bridgmanite under lower mantle conditions. Considering this fact, the spin crossover of [Fe$^{3+}$]$_{Si}$ in bridgmanite may not have the same noticeable impact on lower mantle properties as in (Mg,Fe)O ferropericlase  \citep{Wu14}. Nevertheless, the volume reduction and its associated elastic anomalies due to HS and LS crossover of [Fe$^{3+}$]$_{Si}$ should be taken into account  in calculations of thermodynamic equilibrium in the lower mantle.    

\section{Conclusions}
We have presented a first-principles LDA + U$_{sc}$ investigation of the spin crossover in (Mg,Fe$^{3+}$)(Si,Fe$^{3+}$)O$_3$ bridgmanite.  In order to investigate the system close to experimental conditions,  a disordered substitution of Fe$^{3+}$ was modeled using complete sampling of configurations in super-cells containing up to 80 atoms. Thermal effects were properly captured by calculating the vibrational contribution to the free-energy within the quasi harmonic approximation.  Disorder of Fe$^{3+}$ does not seem to affect the HS to LS crossover of [Fe$^{3+}$]$_{Si}$ significantly. The crossover pressure increases with increasing Fe$^{3+}$ concentration, which is consistent with experimental observations \citep{Lin12,Mao15}. Comparison of our calculated compression curves with that of \citet{Mao15} suggests that in a closed system, Fe$^{2+}$ present in the sample may tend to transform into Fe$^{3+}$ by introduction of Mg and O vacancies with increasing pressure, whose arrangements is still unclear. The spin crossover is also accompanied by a significant volume reduction and by an anomalous decrease in the bulk modulus across the crossover region, in contrast with previous calculations. Further investigations of the effect of the spin crossover on elastic properties at lower mantle conditions are needed to better understand the thermal structure and the composition of the Earth's lower mantle. The present study establishes a foundation for such study. 

\appendix
\section{Solid solution model for Br10 sample}
Here, we describe the Br10 model which has been used to analyze the compression curve obtained by \textit{Mao et al.}, 2015 for the bridgmanite sample (Br10) containing Fe$^{3+}$/$\sum Fe$ $\approx$ 20\% in the octahedral sites (Si-sites) and Fe$^{2+}$/$\sum Fe$ $\approx$ 80\% in the Mg-sites.  We model Br10 by balancing the stoichiometry with appropriate molar fractions of (Mg$_{1-x}$Fe$^{2+}_{x}$)SiO$_3$ and (Mg$_{1-x}$Fe$^{3+}_{x}$)(Si$_{1-x}$Fe$^{3+}_{x}$)O$_3$ bridgmanite. We vary Fe$^{3+}_{Si}$/$\sum$Fe from 10\% to 20\% to assess the effect of Fe$^{3+}$ variation on the compression curve. The volume of Br10 is modeled using an ideal solid solution between Fe$^{2+}$- and Fe$^{3+}$-bearing bridgmanite.

The stoichiometry for the sample containing 10\% Fe$^{3+}_{Si}$/$\sum$Fe (lower bound) is obtained as follows:
\begin{eqnarray}
(Mg_{0.9}Fe_{0.1})SiO_3 &\rightarrow&  (Mg_{0.92}Fe^{2+}_{0.08})SiO_3 + (Mg_{0.99}Fe^{3+}_{0.01})(Si_{0.99}Fe^{3+}_{0.01})O_3 \nonumber \\
&-& 0.99MgSiO_3 - 0.02MgO - 0.005O_2.
\end{eqnarray}
Transferring equal amount of MgSiO$_3$ to both Fe$^{2+}$- and Fe$^{3+}$-bearing MgSiO$_3$, we obtain
\begin{eqnarray}
(Mg_{0.9}Fe_{0.1})SiO_3 &\rightarrow&  \left[(Mg_{0.92}Fe^{2+}_{0.08})SiO_3 - 0.495MgSiO_3  \right] \nonumber \\ &+& \left[ (Mg_{0.99}Fe^{3+}_{0.01})(Si_{0.99}Fe^{3+}_{0.01})O_3 - 0.495MgSiO_3 \right] \nonumber \\ &-& 0.02MgO - 0.005O_2,
\end{eqnarray}
or
\begin{eqnarray}
(Mg_{0.9}Fe_{0.1})SiO_3 &\rightarrow&  0.505(Mg_{0.84158}Fe^{2+}_{0.15842})SiO_3  \nonumber \\ &+& 0.505(Mg_{0.9802}Fe^{3+}_{0.019802})(Si_{0.9802}Fe^{3+}_{0.019802})O_3   \nonumber \\ &-& 0.02MgO - 0.005O_2.
\end{eqnarray}
We chose to incorporate $MgO$ and $O_2$ as $Mg$ and $O$ vacancies in Fe$^{3+}$-bearing MgSiO$_3$:
\begin{eqnarray}
(Mg_{0.9}Fe_{0.1})SiO_3 &\rightarrow&  0.505(Mg_{0.84158}Fe^{2+}_{0.15842})SiO_3  \nonumber \\ &+& 0.505\left[(Mg_{0.9802}Fe^{3+}_{0.019802})(Si_{0.9802}Fe^{3+}_{0.019802})O_3 \right.  \nonumber \\ 
   &-& \left. 0.039604Mg - 0.059406O\right].
\end{eqnarray}
The above equation can be rearranged as 
\begin{eqnarray} \label{Br10_SI}
 (Mg_{0.9}Fe_{0.1})SiO_3 &\rightarrow& a(Mg_{1-x}Fe^{2+}_{x})SiO_{3} \nonumber \\ &+&
 b(Mg_{1-y - \alpha}Fe^{3+}_{y})(Si_{1-y}Fe^{3+}_{y})O_{3(1-\beta)},
\end{eqnarray}
where Mg and O vacancies are constrained by Fe$^{3+}$ concentration as $\alpha$ = 2$y$ and $\beta$ = $y$. Owing to the fact that the total iron concentration in the sample is fixed, these stoichiometric coefficients are also constrained by $ax + 2by = 0.1$. For the sample containing 10\% [Fe$^{3+}]_{Si}$/$\sum$Fe (lower bound): $a$ = $b$ = 0.505, $x$ = 0.15842, and $y$ = 0.019802. Similarly for 20\% [Fe$^{3+}]_{Si}$/$\sum$Fe (upper bound): $a$ = $b$ = 0.51, $x$ = 0.11765, $y$ = 0.039216.  

It is worth mentioning that the choice of the stoichiometric coefficients $a$ and $b$ in Eq.~\ref{Br10_SI} is not unique. However, since the volume variation with respect to iron substitution in bridgmanite has been assumed to be linear and we are assuming an ideal solid solution of Fe$^{2+}$- and Fe$^{3+}$-bearing bridgmanite, other choices of $a$ and $b$ would provide the same estimate for the compression curve of our B10 model.

\textbf{Acknowledgments}

This work was supported primarily by grants NSF/EAR -1319368, 1348066, and NSF/CAREER 1151738. 
Computations were performed at the Minnesota Supercomputing Institute (MSI) and at the Blue Waters System at NCSA.

\end{document}